# Multi-step stochastic mechanism of polarization reversal in rhombohedral ferroelectrics

Y. A. Genenko[1*], R. Khachaturyan[2], I. S. Vorotiahin[1], J. Schultheiß[3,4], J. E. Daniels[5], A. Grünebohm[2] and J. Koruza[3]

[1]*Department of Materials and Earth Sciences, Technical University of Darmstadt, Otto-Berndt-Str. 3, 64287 Darmstadt, Germany*

[2]*Interdisciplinary Center for Advanced Materials Simulation, Ruhr-Universität Bochum, Universitätsstr. 150, 44801 Bochum, Germany*

[3]*Department of Materials and Earth Sciences, Technical University of Darmstadt, Alarich-Weiss-Straße 2, 64287 Darmstadt, Germany*

[4]*Department of Materials Science and Engineering, NTNU Norwegian University of Science and Technology, 7034 Trondheim, Norway*

[5]*School of Materials Science and Engineering, UNSW Sydney, NSW, 2052, Australia*

## Abstract

A stochastic model for the field-driven polarization reversal in rhombohedral ferroelectrics is developed, providing a description of their temporal electromechanical response. Application of the model to simultaneous measurements of polarization and strain kinetics in a rhombohedral Pb(Zr,Ti)O$_3$ ceramic over a wide time window allows identification of preferable switching paths, fractions of individual switching processes, and their activation fields. Complementary, the phenomenological Landau-Ginzburg-Devonshire theory is used to analyze the impact of external field and stress on switching barriers showing that residual mechanical stress may promote the fast switching.

*Corresponding author: genenko@mm.tu-darmstadt.de

1. Introduction

Ferroelectrics are functional materials used for a wide range of important applications, ranging from actuators and sensors [1] to ferroelectric memories (FERAM) [2]. Relevant physical properties, such as permittivity [3], piezoresponse [4] and fracture toughness [5], strongly depend on the crystallographic structure and can be affected by composition [6,7], temperature, and application of mechanical stresses [8]. Furthermore, the appearance of a spontaneous polarization and the possibility to switch its direction by an electric field are key features for ferroelectric memories [2], especially for multi-bit data storage [9-12]. Thus, the manipulation and optimization of the switching kinetics is of high relevance. However, the understanding of the underlying mechanisms is still incomplete. In particular, the fact that most ferroelectrics are also ferroelastics with the second order parameter spontaneous strain [13], results in complex switching paths.

Macroscopic polarization switching kinetics is commonly described by stochastic models, such as the classical Kolmogorov-Avrami-Ishibashi (KAI) model originally developed to describe melt solidification [14]. The KAI model assumes random and statistically-independent nucleation and growth of reversed polarized domains in a uniform polarized medium [15,16]. Thereby individual random switching events are assumed to occur in parallel. Problematically, the KAI model involves only one polarization component, which makes it impossible to describe the electric-field induced switching of the spontaneous strain of ferroelectrics/ferroelastics.

Switching mechanisms in ferroelastic/ferroelectrics are subject to controversial discussion [17-24], and two successive non-180° events [17-19,22,23], predominantly 180° events [21], or exclusively single 180° events [24] have been predicted. Experimentally, consecutive 90°- or, generally, non-180°-switching events have been observed by *in situ* x-ray diffraction measurements [17] and ultrasonic investigations [18] and two characteristic times for sequential polarization reorientation have been identified [19]. Some reports suggest that 180°- and non-180°switching events might occur in parallel [20,21].

To get insight in the complicated switching mechanisms, the macroscopic strain of the ferroelectrics/ferroelastics can be measured simultaneously with the switched polarization. Recently such measurements supported by *in situ* X-ray diffraction experiments were performed on a polycrystalline ferroelectric lead zirconate titanate (PZT) with tetragonal symmetry [25]. Based on these experimental data, an original multistep stochastic mechanism (MSM) model was suggested [26], which allows describing the simultaneous polarization and strain responses of tetragonal ferroelectrics over a broad time range with high accuracy [27]. Particularly, the MSM model allows one to resolve the fraction of ferroelastically-active 90°-switching events. However, in ferroelectrics with rhombohedral and orthorhombic crystallographic symmetries, also 109°-, 71°-, 120°- and 60°-switching events become possible for which no stochastic model is available at the moment. Since in particular rhombohedral PZT has good prospects for single-phase electromechanical applications due to its low coercive fields, largest strain responses in thin film geometry [7], and fast switching [28,29], this is a severe shortcoming. Only a few other models for analysis of the electromechanical response of rhombohedral ferroelectrics have been advanced [30-33], which, however, do not account for the time-resolved response of strain and polarization. Chen and Lynch [30] studied the quasi-static field dependences of polarization and stress, using a computational micromechanics model. A combined switching assumption (CSA) model suggested by Li and Rajapakse, assuming that two or three types of non-180° switching occur simultaneously, exhibited good qualitative agreement with experimental polarization-field and stress-strain loops, but had no access to kinetics [31]. The analysis by Hall *et al*. demonstrated that stresses due to intergranular coupling play an important role during polarization reversal in polycrystalline rhombohedral ferroelectric/ferroelastic PZT ceramics [32]. Various domain structures were also studied in rhombohedral barium titanate by means of molecular dynamics, but only in statics at zero-temperature conditions, thus excluding polarization switching processes [33].

The current work is devoted to the development of a stochastic model for the kinetics of multistep polarization switching processes in rhombohedral ferroelectrics/ferroelastics. The paper is organized as follows: In Section 2, the complexity of extending the MSM model to rhombohedral structures is explained and the possible switching paths are introduced. The classical KAI model is then extended by including all possible sequential non-180° polarization reorientation steps and parallel 180° switching events. To comprehend the simultaneous strain kinetics, a relation between the time-dependent strain and polarization is derived. In Section 3, the likelihood of the switching paths is evaluated by means of the phenomenological Landau-Ginzburg-Devonshire (LGD) model. In Section 4, the above models are applied to original polarization and strain switching experiments over a time domain from $10^{-6}$ s to $10^{1}$ s for a range of applied electric field values around the coercive field of the sample, and the fitting results are discussed based on the concepts from Sections 2 and 3. Finally, the results are concluded in Section 5.

## 2. Theory of consecutive stochastic polarization switching processes

In a rhombohedral ferroelectric, local polarization may adopt one of the eight possible directions along the body diagonals of the pseudo-cubic cell (<111> directions). Application of an electric field opposite to the polarization direction induces reversal of the polarization, which may consist of different intermediate switching events. In this Section, we first define and display possible switching paths and then calculate their probabilities by an extension of the KAI approach. Finally, we derive the variation of strain according to the polarization variation.

## A. Polarization rotations and switching channels

Consider a crystalline unit cell of a rhombohedral ferroelectric. Axes of a Cartesian frame are chosen to be collinear with main axes of a pseudo-cubic cell (Fig. 1(a)). The system is assumed to be initially in the polarization state $P_s(-1,-1,-1)/\sqrt{3}$ with spontaneous polarization $P_s$ and then it is driven to the final state $P_s(1,1,1)/\sqrt{3}$ by application of the electric field **E** pointing into the $[111]$ direction. Polarization reversal may proceed along different paths exemplarily shown in Fig. 1(a), namely, by a direct reversal path A-D, by two consecutive 109°- and 71°- polarization rotations A-E-D, by two consecutive 71°- and 109°-polarization rotations A-B-D and by a triple consecutive 71°- polarization rotation A-B-C-D. It should be noted that here the term "rotation" is not used in the same sense as in monoclinic ferroelectrics where the polar vector can freely rotate in given crystallographic plane, but instead refers to a rotation path through a sequence of fixed polarization directions.

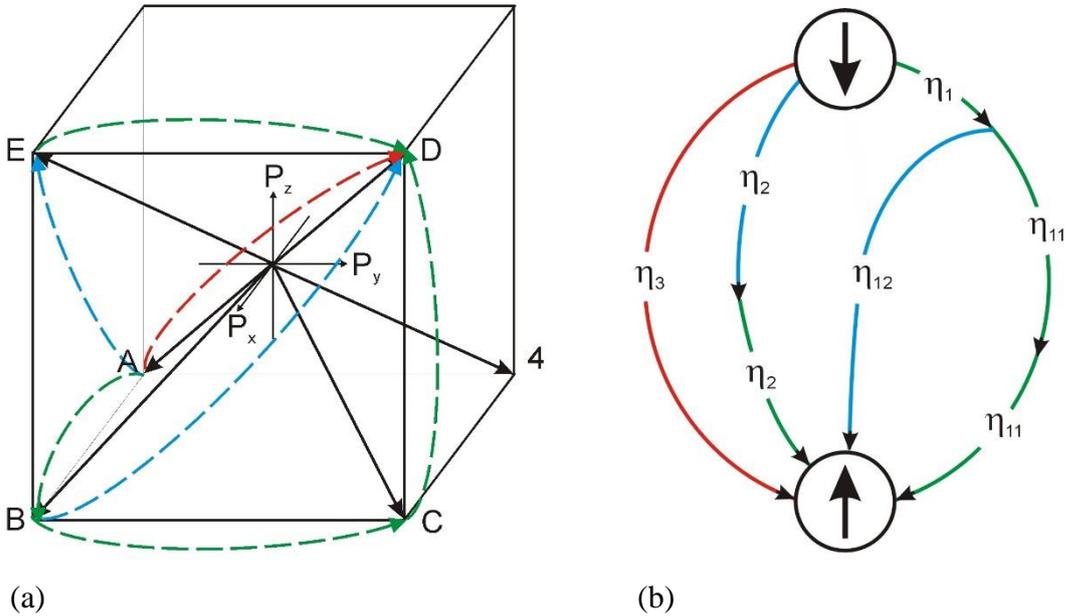

(a) (b)

Fig. 1. (a) Possible realizations of the field-driven polarization reversal in rhombohedral state: exemplary 71°-71°-71° polarization rotation path A-B-C-D, exemplary 71°-109° polarization rotation path A-B-D, exemplary 109°-71° polarization rotation path A-E-D and a direct 180° polarization reversal A-D. (b) Definition of fractions of different polarizations switching paths. Both in (a) and (b) 71°-switching processes are indicated by green lines, 109°-switching processes by blue lines and 180°-switching processes by red lines. Black arrows indicate the end of each process.

We define the probabilities for the first switching event as $\eta_1$, $\eta_2$ and $\eta_3$ for switching starting with 71°, 109° and 180° rotations, respectively, and exclude all other more complex switching paths, i.e. $\eta_1+\eta_2+\eta_3 =1$. After the first polarization rotation, the first switching channel splits up in two possible paths, single 109°-polarization rotation with a weight $\eta_{12}$ and double 71°-polarization rotation with weight $\eta_{11}$, satisfying $\eta_{11}+\eta_{12} = \eta_1$. In a ferroelectric, switching probabilities may depend on energy barriers between different polarization states, strength of applied field and/or mechanical stress, as well as electric and elastic interactions between domains and boundary conditions.

### B. Extension of the KAI model to consecutive switching events

In the classical KAI approach [14-16], switching processes at all different locations and times are considered as statistically independent. This cannot be true when accounting for consecutive switching events, whereby some polarization steps may only occur after previous ones are finished. The assumption of statistical independence of switching events at different locations is also questionable. In fact, piezoresponse force microscopy and transmission electron microscopy revealed clustering during polarization reversal ranging from a few grains [34] to agglomerations of $10^2$–$10^3$ grains [35,36] in polycrystalline thin films. On the other hand, the grain-resolved three-dimensional X-ray diffraction in bulk ceramics disclosed a collective dynamics correlated over about 10–20 grains [37-39]. Analysis of polarization and field correlations, based on the self-consistent mesoscopic switching model [40-42], revealed that the electric field-mediated correlations in bulk ceramics remain short- range at the typical scale of the grain size. Physical background of the short-range correlations is an effective screening of depolarization fields by adapting surface charges at grain boundaries. The question of correlations due to elastic interactions, however, remains open [25]. We retain here the simplified KAI assumption that switching processes at different locations are statistically independent, thus neglecting the electric and elastic interactions of domains during their switching, and focus on the sequential switching steps. Following the MSM approach [26], we derive the probabilities of different consecutive switching events.

Initially, we assume a monodomain single crystal with saturation polarization along $\left[\bar{1}\,\bar{1}\,\bar{1}\right]$. When an opposite field is applied, the local polarization may experience different switching events with different nucleation rates per unit time and unit volume and exhibit different geometries of growing switched domains. We first consider, in the spirit of the KAI model [14-16], the nucleation of reversed domains according to the single 180°-switching process A-D (Fig. 1(a)) by the application of electric field pulse starting at time $t_0 = 0$. The key conceptual element of this approach, introduced by Ishibashi and Takagi [16], is the probability $q_{180}(t,t_0)$ for an arbitrary point not to be comprised by a switched area of some domain,

$$q_{180}(t,t_0) = \exp\left[-\left(\frac{t-t_0}{\tau_3}\right)^{\beta_3}\right] \qquad (1)$$

where $\tau_3$ is the characteristic time of 180°-switching process and $\beta_3$ is the so called Avrami index related to the dimensionality $D$ of the growing domain. Ishibashi and Takagi [16] considered two alternative regimes of the reversed domain nucleation: (I) the nucleation rate is constant in space and time throughout the switching period when the constant uniform field is kept applied, and (II) there are only latent nuclei but no new nucleation. Under the prerequisite that the velocity of domain growth is constant (though field dependent), $\beta_3 = D+1$ applies for the regime I, and thus $\beta_3$ is larger than unity, while for the regime II, $\beta_3 = D$. This provides a contribution to the total switched polarization proportional to the fraction $(1-q_{180}(t,t_0))$ of the system volume and reads

$$\Delta p_{180}(t) = \eta_3 2P_s \left\{ 1 - \exp\left[ -\left(\frac{t}{\tau_3}\right)^{\beta_3} \right] \right\} \quad (2)$$

The MSM model extended the KAI approach by application of the formula (1) to the first and the second switching events of the two-step switching processes with their specific characteristic times and Avrami indices. Similarly to the 90°-90° polarization rotation sequence in tetragonal ferroelectrics [26], for the 109°-71° polarization rotation path A-E-D (Fig. 1(a)) in rhombohedral counterparts, we introduce a probability $q_{109}(t_1, t_0)$ not to switch according to the first switching mechanism by some intermediate time $t_1$. Further, we introduce a probability $q_{71}(t, t_1)$ not to switch according to the second switching mechanism by the time $t$ after the first switching event occurred at time $t_1$. When summarizing over all possible intermediate times $t_1$ between 0 and $t$, the total probability to switch once according to the first mechanism and not to switch anymore according to the second mechanism by the time $t$ is obtained,

$$L_{20}(t) = \frac{\beta_2}{\tau_2} \int_0^t dt_1 \left(\frac{t_1}{\tau_2}\right)^{\beta_2 - 1} \exp\left[-\left(\frac{t_1}{\tau_2}\right)^{\beta_2}\right] \exp\left[-\left(\frac{t-t_1}{\tau_{21}}\right)^{\beta_{21}}\right] \quad (3)$$

where $\tau_2$ and $\tau_{21}$ are the characteristic times of the first 109°- and the second 71°-processes, respectively, and $\beta_2$ and $\beta_{21}$ are the respective Avrami indices. Accordingly, the total probability to switch firstly according to the first mechanism and secondly according to the second mechanism by the time $t$ reads as

$$\begin{aligned} L_{21}(t) &= \frac{\beta_2}{\tau_2} \int_0^t dt_1 \left(\frac{t_1}{\tau_2}\right)^{\beta_2 - 1} \exp\left[-\left(\frac{t_1}{\tau_2}\right)^{\beta_2}\right] \left\{ 1 - \exp\left[-\left(\frac{t-t_1}{\tau_{21}}\right)^{\beta_{21}}\right] \right\} \\ &= 1 - \exp\left[-\left(\frac{t_1}{\tau_2}\right)^{\beta_2}\right] - L_{20}(t). \end{aligned} \quad (4)$$

This provides a contribution to the total switched polarization

$$\Delta P_{109-71}(t) = \eta_2 \left[ \frac{4}{3} P_s L_{20}(t) + 2 P_s L_{21}(t) \right] \tag{5}$$

since each 109°-polarization rotation provides variation by 4P$_s$/3 and each 71°-polarization rotation provides variation by 2P$_s$/3 along the $[111]$ direction (see Fig. 1(a)).

In the case of the first 71°-switching event, the description of polarization reversal becomes more complicated because it allows splitting in two channels as depicted in Fig. 1(b). The probabilities for the two-step 71°-109° channel are formally similar to the above considered 109°-71° process and can be described by formulas similar to Eqs. (3,4). Namely, the total probability to switch once according to the first 71°-mechanism and not to switch anymore by the second 109°-rotation reads

$$L_{102}(t) = \frac{\beta_1}{\tau_1} \int_0^t dt_1 \left( \frac{t_1}{\tau_1} \right)^{\beta_1-1} \exp\left[ -\left( \frac{t_1}{\tau_1} \right)^{\beta_1} \right] \exp\left[ -\left( \frac{t-t_1}{\tau_{12}} \right)^{\beta_{12}} \right], \tag{6}$$

where $\tau_1$ and $\tau_{12}$ are the characteristic times of the first 71°-and the second 109°- processes, respectively, and $\beta_1$ and $\beta_{12}$ are the respective Avrami indices. The total probability to switch firstly by 71° and secondly by 109° by the time $t$ reads as

$$L_{12}(t) = \frac{\beta_1}{\tau_1} \int_0^t dt_1 \left( \frac{t_1}{\tau_1} \right)^{\beta_1-1} \exp\left[ -\left( \frac{t_1}{\tau_1} \right)^{\beta_1} \right] \left\{ 1 - \exp\left[ -\left( \frac{t-t_1}{\tau_{12}} \right)^{\beta_{12}} \right] \right\}$$
$$= 1 - \exp\left[ -\left( \frac{t_1}{\tau_1} \right)^{\beta_1} \right] - L_{102}(t). \tag{7}$$

We note that the characteristic times and Avrami indices may differ between 71°-109° and 109°-71° switching since the energy barrier for both rotation angles depends on initial polarization and strain configuration. The 71°-109° switching path provides a contribution to the total switched polarization

$$\Delta P_{71-109}(t) = \eta_{12}\left[\frac{2}{3}P_s L_{102}(t) + 2P_s L_{12}(t)\right]. \tag{8}$$

A succession of three 71°- switching processes (such as the A-B-C-D path in Fig. 1(a)) can be considered in a similar way with intermediate times $t_1$ and $t_2$ for the second and the third event, respectively. When integrating over all possible intermediate times, the probability to switch first time on the 71°-71°-71° path and not to switch anymore reads

$$L_{101}(t) = \frac{\beta_1}{\tau_1}\int_0^t dt_1 \left(\frac{t_1}{\tau_1}\right)^{\beta_1 - 1} \exp\left[-\left(\frac{t_1}{\tau_1}\right)^{\beta_1}\right] \exp\left[-\left(\frac{t - t_1}{\tau_{11}}\right)^{\beta_{11}}\right], \tag{9}$$

where $\tau_1$ and $\tau_{11}$ are the characteristic times of the first 71°- and the second 71°-processes, respectively, and $\beta_1$ and $\beta_{11}$ are the respective Avrami indices. Similarly, the probability to switch the first and the second time on the 71°-71°-71° path and not to switch anymore is

$$L_{110}(t) = \frac{\beta_1}{\tau_1}\int_0^t dt_1 \left(\frac{t_1}{\tau_1}\right)^{\beta_1 - 1} \exp\left[-\left(\frac{t_1}{\tau_1}\right)^{\beta_1}\right] \int_{t_1}^t dt_2 \frac{\beta_{11}}{\tau_{11}}\left(\frac{t_2 - t_1}{\tau_{11}}\right)^{\beta_{11} - 1} \exp\left[-\left(\frac{t_2 - t_1}{\tau_{11}}\right)^{\beta_{11}}\right]$$
$$\times \exp\left[-\left(\frac{t - t_2}{\tau_{111}}\right)^{\beta_{111}}\right] \tag{10}$$

where $\tau_{111}$ and $\beta_{111}$ are the characteristic time and the respective Avrami index of the third 71°-switching event. Finally, the probability to consequently switch three times by 71° equals

$$L_{111}(t) = \frac{\beta_1}{\tau_1}\int_0^t dt_1 \left(\frac{t_1}{\tau_1}\right)^{\beta_1 - 1} \exp\left[-\left(\frac{t_1}{\tau_1}\right)^{\beta_1}\right] \int_{t_1}^t dt_2 \frac{\beta_{11}}{\tau_{11}}\left(\frac{t_2 - t_1}{\tau_{11}}\right)^{\beta_{11} - 1} \exp\left[-\left(\frac{t_2 - t_1}{\tau_{11}}\right)^{\beta_{11}}\right]$$
$$\times \left\{1 - \exp\left[-\left(\frac{t - t_2}{\tau_{111}}\right)^{\beta_{111}}\right]\right\}. \tag{11}$$

The contribution to the total switched polarization for the path 71°-71°-71° is thus

$$\Delta P_{71-71-71}(t) = \eta_{11}\left[\frac{2}{3}P_s L_{101}(t) + \frac{4}{3}P_s L_{110}(t) + 2P_s L_{111}(t)\right]. \tag{12}$$

Taken together, the contributions of all switching channels sum up to a total polarization variation of

$$\Delta P(t) = \eta_3 2P_s \left\{ 1 - \exp\left[ -\left(\frac{t}{\tau_3}\right)^{\beta_3} \right] \right\} + \eta_2 \left[ \frac{4}{3} P_s L_{20}(t) + 2P_s L_{21}(t) \right] + \eta_{12} \left[ \frac{2}{3} P_s L_{102}(t) + 2P_s L_{12}(t) \right]$$
$$+ \eta_{11} \left[ \frac{2}{3} P_s L_{101}(t) + \frac{4}{3} P_s L_{110}(t) + 2P_s L_{111}(t) \right]. \quad (13)$$

We note that the time dependences of probabilities $L_i(t)$ do not mean a gradual transformation of each particular ferroelectric cell. In line with the KAI concept, these probabilities denote the parts of the system, covered with the respective switched domains. The local switching at a certain place is assumed to occur instantaneously as soon as this location is comprised by a switched area of a growing domain.

Being a direct extension of the KAI approach to the case of double and triple switching events, the formula (13) includes additional integrations over intermediate time steps and cannot be solved in a closed form. In order to illustrate qualitative trends, it is convenient to set all beta values to unity. In this case the involved event probabilities become

$$L_{20}(t) = \frac{\tau_{21}}{\tau_2 - \tau_{21}} \left( e^{-t/\tau_2} - e^{-t/\tau_{21}} \right), \quad (14a)$$

$$L_{21}(t) = 1 - e^{-t/\tau_2} - \frac{\tau_{21}}{\tau_2 - \tau_{21}} \left( e^{-t/\tau_2} - e^{-t/\tau_{21}} \right), \quad (14b)$$

$$L_{102}(t) = \frac{\tau_{12}}{\tau_1 - \tau_{12}} \left( e^{-t/\tau_1} - e^{-t/\tau_{12}} \right), \quad (14c)$$

$$L_{12}(t) = 1 - e^{-t/\tau_1} - \frac{\tau_{12}}{\tau_1 - \tau_{12}} \left( e^{-t/\tau_1} - e^{-t/\tau_{12}} \right), \quad (14d)$$

$$L_{101}(t) = \frac{\tau_{11}}{\tau_1 - \tau_{11}} \left( e^{-t/\tau_1} - e^{-t/\tau_{11}} \right), \quad (14e)$$

$$L_{110}(t) = \frac{\tau_{111}}{\tau_{11} - \tau_{111}} \left[ \frac{\tau_{11}}{\tau_{11} - \tau_1} \left( e^{-t/\tau_{11}} - e^{-t/\tau_1} \right) - \frac{\tau_{111}}{\tau_{111} - \tau_1} \left( e^{-t/\tau_{111}} - e^{-t/\tau_1} \right) \right], \quad (14f)$$

$$L_{111}(t) = 1 - e^{-t/\tau_1} - \frac{\tau_{11}^2}{(\tau_{11} - \tau_{111})(\tau_{11} - \tau_1)}\left(e^{-t/\tau_{11}} - e^{-t/\tau_1}\right) + \frac{\tau_{111}^2}{(\tau_{11} - \tau_{111})(\tau_{111} - \tau_1)}\left(e^{-t/\tau_{111}} - e^{-t/\tau_1}\right). \quad (14g)$$

In the following, we do not use any assumptions for beta or the other fitting parameters and use Eq. (13) together with expressions (3,4,6,7,9-11) to describe experimental data.

### C. Relation between polarization and strain

To establish a relation between the macroscopic strain, $S$, and the macroscopic polarization, $p$, during the polarization reversal a general relation can be used [43,44],

$$S_{ij} = Q_{ijmn} p_m p_n \quad (15)$$

with the electrostriction tensor $Q_{ijmn}$ and summation over the repeated indices implied, if no stress is applied to the system. For ferroelectrics with a cubic parent phase, the piezoelectric contribution can be derived from Eq. (15) when the spontaneous polarization $P$ is singled out as

$$p_n \cong P_n + \varepsilon_0 \varepsilon_{nm} E_m \quad (16)$$

with the permittivity of vacuum $\varepsilon_0$ and the relative permittivity of the ferroelectric $\varepsilon_{nm}$. By substitution of Eq. (16) into Eq. (15) one obtains

$$S_{ij} \cong Q_{ijmn} P_m P_n + d_{ijk} E_k + \varepsilon_o^2 Q_{ijmn} \varepsilon_{ml} \varepsilon_{nk} E_l E_k \quad (17)$$

and an expression for the piezoelectric coefficient [43]

$$d_{ijk} = 2\varepsilon_0 \varepsilon_{km} Q_{ijml} P_l \,. \quad (18)$$

The second, piezoelectric term in Eq. (17), linear in electric field, amounts at maximum about 0.1 of the spontaneous polarization contribution at highest field values. The last term in Eq.

(17), quadratic in electric field, amounts at maximum about $3\times10^{-4}$ of the spontaneous polarization contribution at highest field values and therefore will be neglected in the following.

Experimental measurements of polarization and strain performed on a polycrystalline ferroelectric in the direction of the applied field will be approximated by the presented single crystal model with the field applied in the [111]-direction. Nevertheless, it makes sense to specify the relations (15,17,18) in the "parent" coordinate frame of Fig. 1(a) collinear with the cubic parent phase, where the tensor of electrostriction has a particularly simple form [44]. This way all coefficients depending on electrostriction can be expressed by the matrix elements $Q_{11}, Q_{12}$, and $Q_{44}$ [44] (Voigt notation). The strain variation along the field direction $\Delta S'_{33}(t)$ in the new coordinate frame $(x', y', z')$, such that the $z'$ axis points in the field direction, may then be determined as [13]

$$\Delta \mathbf{S}' = \mathbf{T}\Delta \mathbf{S}\mathbf{T}^{-1} \text{ with } \mathbf{T} = \begin{pmatrix} 0 & 1/\sqrt{2} & -1/\sqrt{2} \\ -2/\sqrt{6} & 1/\sqrt{6} & 1/\sqrt{6} \\ 1/\sqrt{3} & 1/\sqrt{3} & 1/\sqrt{3} \end{pmatrix}. \tag{19}$$

This reveals the time dependent strain variation along the field direction

$$\Delta S(t) = \Delta S'_S(t) + \Delta S'_P(t), \tag{20}$$

where the first term presents a contribution of the spontaneous polarization variation to the strain (the first term in Eq. (17)) and the second one a contribution of the piezoelectric effect (the second term in Eq. (17)). Since each 71°- and each 109°- rotation of the polarization of the amplitude $P_s$ contributes the maximum strain variation,

$$\Delta S_{\max} = -\frac{16}{9}Q_{44}P_s^2, \tag{21}$$

the total variation of strain due to the spontaneous polarization equals

$$\Delta S_S^{'}(t) = \Delta S_{\max} \left[ \eta_2 L_{20}(t) + \eta_{12} L_{102}(t) + \eta_{11} \left( L_{101}(t) + L_{110}(t) \right) \right]. \tag{22}$$

According to Eq. (17) the piezoelectric strain is coupled to the spontaneous polarization by the piezoelectric coefficient (18), proportional itself to the spontaneous polarization, that results in

$$\Delta S_P^{'}(t) = \frac{2\varepsilon_0 \varepsilon_c}{3} \left( Q_{11} + 2Q_{12} + 4Q_{44} \right) \left( \Delta P(t) - P_s \right) E, \tag{23}$$

where $E$ is the electric field applied in the $z'$ direction, $\varepsilon_c = \varepsilon'_{33}$ is the permittivity in the new coordinate frame and $\Delta P(t)$ is the time dependent polarization variation along the applied field direction given by Eq. (13).

The 180°-switching processes only change the strain by the second term in Eq. (20), which is linear in polarization. In contrast, the non-180° switching events rotate polarization by either 71° or 109°, thus contributing to the variation of the strain by both terms in Eq. (20). It is also important to note that consequent rotations along all paths cause no variation of the strain in the final state by the term quadratic in polarization, Eq. (22), but by the term linear in polarization, Eq. (23). Notice that both formulas for polarization, (13), and strain, (20), present averaging over the whole system and neglect electric and elastic interactions [45] between different switching regions during the polarization reversal, as discussed above in section 2B.

### 3. Model evaluation of polarization switching barriers

In order to interpret the switching rates, obtained by application of our model to experiments, the Landau-Ginzburg-Devonshire approach is used, which allows estimation of barriers for different switching paths. The natural variables of the chosen thermodynamic potential are the spontaneous polarization **P** as an order parameter for the electric subsystem and the strain **S** as an order parameter for the mechanical subsystem. Using a simplified approach for estimations, the expression for the free energy density consists of the Landau and elastic parts (as $\Delta\Phi = G_{Land} + F_{elast}$) [46]:

$$\Delta\Phi = \alpha_{ik}P_iP_k + \beta_{ijkl}P_iP_jP_kP_l + \gamma_{ijklmn}P_iP_jP_kP_lP_mP_n - P_iE_i + \frac{1}{2}c_{ijkl}S_{ij}S_{kl} - \frac{1}{2}q_{ijkl}S_{ij}P_kP_l,$$
(24)

where $\alpha_{ik}$, $\beta_{ijkl}$ and $\gamma_{ijklmn}$ are the Landau expansion coefficients describing the properties of the ferroelectric phase at room temperature, $E_i$ is a component of the applied electric field, $c_{ijkl}$ is the stiffness tensor, $q_{ijkl}$ is the electrostrictive constant tensor. Since gradient terms are omitted in Eq. (24), it describes a monodomain single crystal. Values of the aforementioned coefficients were taken from [47-49] for the PZT 60/40 composition; the applied electric field was considered as variable parameter.

To estimate the role of residual stresses in the polarization reversal we first evaluate the spontaneous strain caused by spontaneous polarization under a stress-free condition. The elastic equation of state derived from the thermodynamic potential (24) reads $\sigma_{ij} = \partial\Delta\Phi/\partial S_{ij}$ with the stress $\sigma_{ij}$, which results in generalized Hooke's law [46]

$$\sigma_{ij} = c_{ijkl}S_{kl} - \frac{1}{2}q_{ijkl}P_kP_l .$$
(25)

The spontaneous strain is defined by the condition $\sigma_{ij} = 0$ and reads

$$S_{ij}^0 = \frac{1}{2}s_{ijmn}q_{mnkl}P_kP_l = Q_{ijkl}P_kP_l$$
(26)

with the elastic compliance tensor $s_{ijmn}$. The spontaneous polarization at zero stress is then obtained by substituting the spontaneous strain (26) in (24) and minimizing the thermodynamic potential using the condition $\partial\Delta\Phi/\partial P_i = 0$ [46].

When considering a polycrystalline material, we assume that the spontaneous strain of each grain cannot be fully realized being resisted by the surrounding [45]. For estimation of this effect on the switching barriers, we assume that the strain at the spontaneous polarization takes a reduced exemplary value $S_{ij}^r = 0.1 S_{ij}^0$, roughly corresponding to the strain reduction in polycrystalline material compared with that of a strain free single crystal [50]. Differently from the stress-free case, the residual stress in this state is not zero anymore and can be evaluated from Eq. (25). By substituting the reduced strain $S_{ij}^r$ in Eq. (24), its effect on polarization states and energy barriers between them may be observed. In the following, we will consider the cases with the full strain (stress-free single-crystal) and the reduced strain (polycrystalline material) in the presence or absence of electric field.

A series of polarization-dependent double-well potentials is obtained from the evaluation of the free energy density with minima located at amplitudes $\pm P_i^r = \pm 0.295$ C/m² when external field is absent and strain is fully developed (Fig. 2(a)). Those values of the polarization amplitudes are applicable to each polarization component. We calculate the energy barriers for 71°, 109° and 180° switching, starting from the state of $P_1 = -P_1^r, P_2 = -P_2^r, P_3 = -P_3^r$. These barriers can determine which path the system will choose from the beginning. In the absence of external electric field and with the full strain, we find the smallest barrier for 71° switching on path A-B in Fig. 1(a), followed by 109°-switching along the path A-E and 180°-switching along the path A-D, which is by factors of 9.98 and 2.99 higher than the first and the second barriers, respectively. If an electric field is applied in the [111] direction, the linear *P-E* coupling in (24) distorts the energy wells and stabilizes the state with $+P_i^r$ (Fig. 2(b)). Since no domain nucleation is considered, the coercive electric fields in the classical Landau theory, i.e. the field values necessary to switch polarization, are an order of magnitude higher than the ones observed experimentally [51]. However, the overall tendency can be anticipated. Most important is that the energy barriers for polarization rotation are modified differently by an external field for the considered switching paths, see Fig. 3(a). The change of the barrier is largest for 180° switching and smallest for 71° switching as the polarization variation is either at minimum (zero) or maximum angle with respect to the field direction. Thus, although the 71°-leap has initially the lowest potential barrier, also its coupling to the applied field is the weakest. When the applied field reaches 10.39 kV/mm (Fig. 2(b)) the barrier for the 71°-switching disappears, making it the most favourable one among the first switching steps. In the same fashion the potential barrier vanishes for the 109°-switching at the applied field of 17.32 kV/mm. As the last one, the barrier for the first 180°-switching disappears at 36.37 kV/mm, making this switching path the least likely of all.

The situation may change if we account for residual stresses, which are introduced into polycrystalline system through intergranular coupling by setting the polarization to the $[\bar{1}\bar{1}\bar{1}]$ direction through electrical poling [45,52,53]. In our LGD model this corresponds to the introduction of the reduced strain $S_{ij}^r$. For *E*=0, the impact of the residual stress on the energy landscape (Eq. (24)) is illustrated in Fig. 2(c). The energy profiles for 180°-switching and 109°-switching paths are hardly changed, while the energy barrier for 71°-switching is enlarged. However, the residual stress modifies the coupling of the switching paths to the field as is seen from a comparison of Fig. 3(a) and 3(b). For the fully developed strain, the switching barrier for 71° switching vanishes as the first one with the field increase (cf. Figs. 2(b) and 3(a)). In contrast, for $S_{ij}^r = 0.1 S_{ij}^0$ the switching barrier for 109° vanishes first. Thus, the chosen residual stress promotes the first 109° switching step at the field value of 17.32 kV/mm as is seen in Fig. 2(d). The barriers for 71°- and 180°-switching disappear consequently at applied fields of 24.25 kV/mm and 34.64 kV/mm, respectively (Fig. 3(b)). Concluding, the priority of first switching steps depends on the residual stress in the poled system and can be in favour of either 71°- or 109°-switching, leaving the 180°-process the least likely scenario.

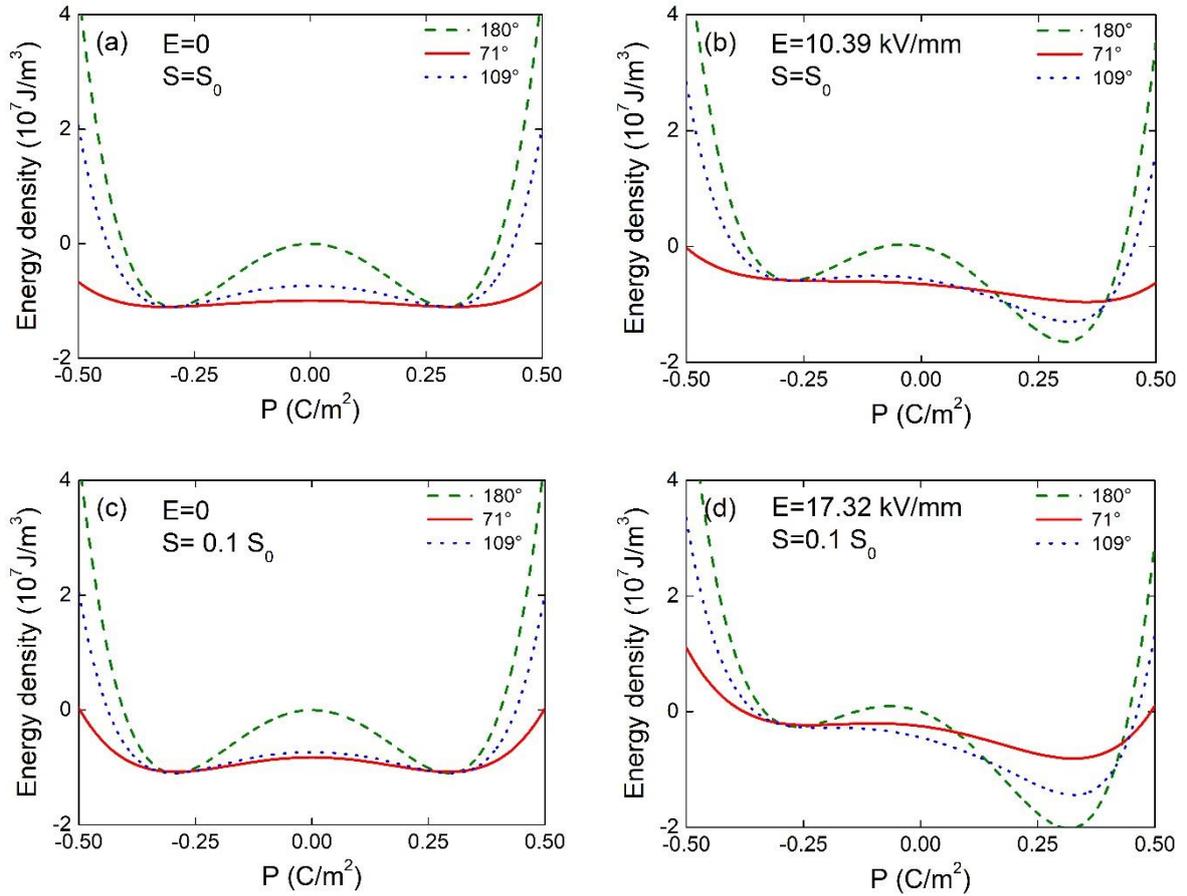

Fig. 2. Free energy density profiles showing potential barriers of the first 71° switching (solid red), the first 109° switching (dotted blue) and the 180° switching (dashed green) processes (a) at the fully developed strain ($S = S_0$) and no applied electric field ($E = 0$) field, (b) at applied electric field $E = 10.39$ kV/mm and strain $S = S_0$, (c) at no applied electric field ($E = 0$) but with a reduced strain equal to $0.1S_0$, and (d) at applied electric field $E = 17.32$ kV/mm and the strain of $0.1S_0$.

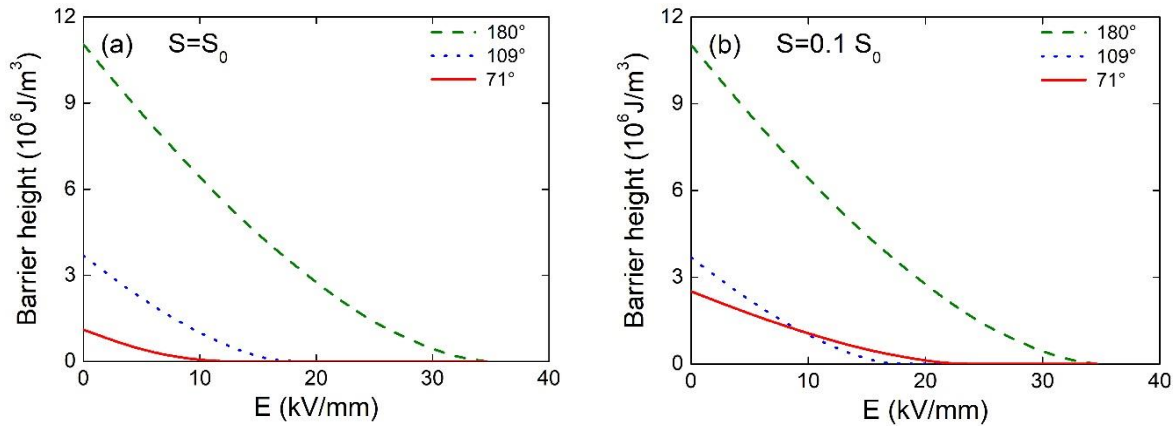

Fig. 3. Field dependence of the energy barrier heights for the first 71° switching (solid red), the first 109° switching (dotted blue) and the 180° switching (dashed green) processes at (a) the fully developed (stress-free) strain ($S = S_0$) and (b) a reduced strain equal to $0.1 S_0$.

## 4. Analysis and discussion of experimental results

*A. Experimental details and sample characterization*

Bulk, polycrystalline $Pb_{0.985}La_{0.01}(Zr_{0.6}Ti_{0.4})O_3$ ceramics of rhombohedral symmetry were prepared by a mixed-oxide route [54]. Standard electromechanical characterization of the sample combining large-signal and small-signal measurements is presented in Fig. 4.

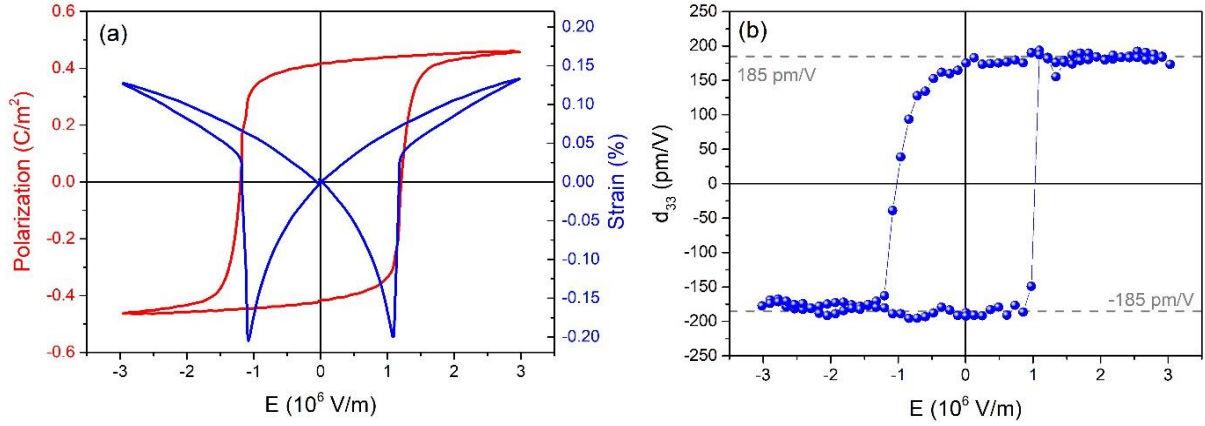

Fig. 4. (a) Bipolar polarization and strain loops (1 Hz). (b) Field-dependent small-signal piezoelectric coefficient in the applied field direction, measured by superimposing a small AC field (500 Hz) over a large DC field (plotted on x-axis).

Moreover, the switched polarization and the macroscopic strain were measured simultaneously as a function of time. The samples were poled in the downward direction with an electric field of 3 kV/mm for 20 s. After a waiting time of 100 s, a 10 s pulse switching field of an amplitude *E* was applied opposite to the poling direction. In order to realize a sharp high voltage (HV) pulse with a rise time of 115 ns, a buffer capacitor, which was charged by a high voltage source (Trek Model 20/20C, Lockport, NY, USA), was combined with a commercial fast HV transistor switch (HTS 41-06-GSM, Behlke GmbH, Kronberg, Germany) [55]. The charge was monitored by measuring the voltage drop across a reference capacitor (WIMA MKS4, Wima, Mannheim, Germany), while the macroscopic displacement of the sample was simultaneously measured by an optical displacement senor (D63, Philtec Inc., Annapolis, MD, USA), which allows a time resolution of $10^{-4}$ s. Further details on the measurement procedures and on the crystallographic and electromechanical characterization of the sample may be found in Refs. [25] and [29], respectively.

## B. Fitting dynamic polarization and strain data

Fig. 5 displays the time-dependent data of polarization and strain for exemplary switching field values $E$ in the range 0.900-1.230 kV/mm for which the switching processes were completed. For the field values outside this range, only initial or final stages of the polarization and strain reversal are observed (displayed in Ref. [29]), which do not allow reliable evaluation of fitting parameters. Note that the leakage current and the reversible dielectric displacement were subtracted in the presented polarization $\Delta P$ data in Fig. 5. The variation of the strain starts in all measurements from zero (poled reference state), which is not explicitly seen in the plot since the data below $10^{-4}$ s are not experimentally accessible.

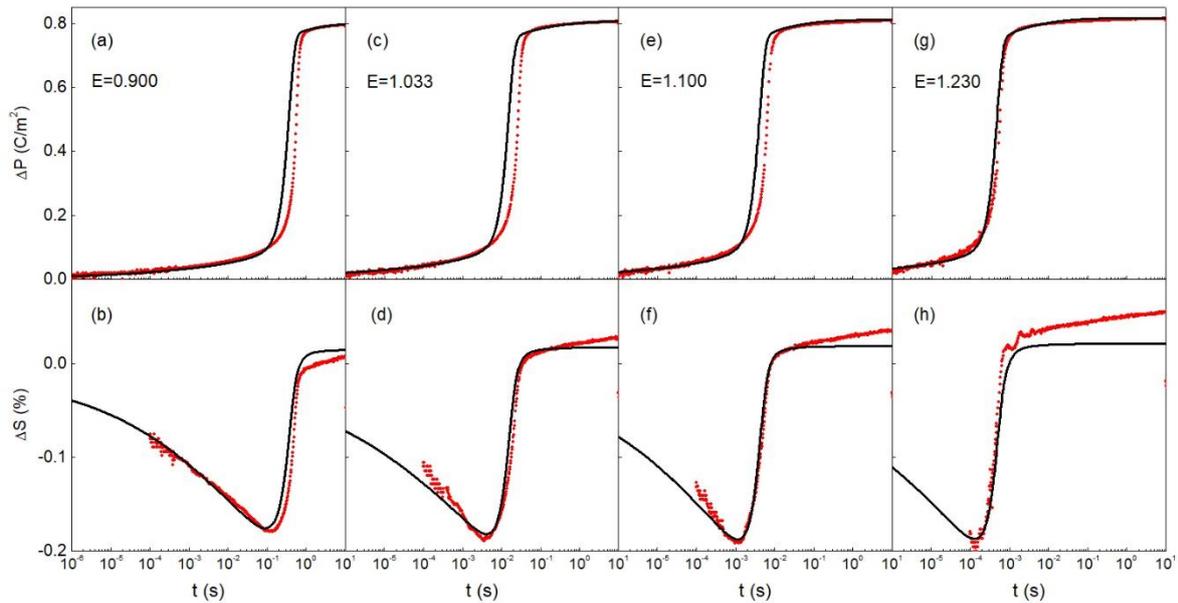

Fig. 5. Variation of the polarization (a,c,e,g) and strain (b,d,f,h) with time during field-induced polarization reversal at different applied field values in kV/mm, as indicated in the plots. Experimental curves are shown by red symbols and fitting curves by black solid lines.

Application of the models for single crystalline materials, presented in Sections 2 and 3, to the polycrystalline material, experimentally studied here, is possible with some reservations. In polycrystalline materials, the orientations of crystalline grains are randomly distributed and the

non-180° domain switching fraction strongly depends on the orientation angle of the grain with respect to the electric field [25]. This leads, on the one hand, to a reduction of the maximum possible polarization of rhombohedral ceramics to the value of $0.866P_s$ [56] and, on the other hand, to the deviation of local polarization directions from directions of local electric fields, which is one of the reasons for the statistical distribution of local switching times [57]. Account of these circumstances allows a much more precise description of polarization kinetics but makes the theory substantially more complicated [27] and is thus not applied here for reasons of transparency. At the same time, application of single crystalline concepts to polycrystalline ceramics of tetragonal symmetry proved to properly capture key features of polarization and strain responses [26]. Hence, in the present study, we restrict ourselves to the direct extension of the KAI model to single crystalline rhombohedral systems subject to a uniform electric field, as presented in Section 2, and recognize its limited ability to describe polycrystalline materials, which can be improved in further work.

For fitting the experimental data in Fig. 5 using Eqs. (13) and (20), some weakly field-dependent material parameters, namely, the total polarization variation $2P_s$, the maximum strain value $\Delta S_{\max}$, and the asymptotic value of the macroscopic piezoelectric coefficient $S_E = 2\varepsilon_0\varepsilon_c(Q_{11}+2Q_{12}+4Q_{44})P_s/3$ in Eq. (23), which can be called a normalized strain, are derived from the experimental data presented in Fig. 4. The fractions of switching processes $\eta_i$, their Avrami indices $\beta_i$ and switching times $\tau_i$ are used as field-dependent fitting parameters. Based on available results for rhombohedral ferroelectrics [29], we implement a similar fitting procedure as developed for tetragonal PZT [26] and assume the Merz law [58] for the field dependence of the switching times: $\tau_i = \tau_0 \exp(E_A^{(i)}/E)$ with the same characteristic time $\tau_0$ and different activation fields $E_A^{(i)}$ for different switching events. The fitting parameters are obtained from a least squares regression with implementation of "A Python Library for Optimizing the Hyperparameters of Machine Learning Algorithms" (Hyperopt) [59], which allows approximation of polarization and strain curves with relative inaccuracy of about 3%. Experimental data are displayed in Fig. 5 by red symbols, while black solid lines highlight the theoretical fitting curves. All parameters used for fitting are presented in Table I. The inaccuracy of the determination of activation fields is below 3%; the inaccuracy of the other field-dependent parameters is characterized by error bars in Fig. 6.

Table I. Parameters used for fitting the experimental curves in Fig. 5 by implementing formulas (13) and (20). Exemplary processes are indicated as in Fig. 1(a). For the field-dependent parameters $\eta_i$ and $\beta_i$ the ranges of their variation with the electric field are displayed.

| Process/ parameter | A-B 1st 71° | B-C 2nd 71° after 71° | C-D 3rd 71° | E-D 2nd 71° after 109° | A-E 1st 109° | B-D 2nd 109° after 71° | A-D 180° |
|---|---|---|---|---|---|---|---|
| $\eta_i$ | 0.28-0.33 | 0.11-0.16 | 0.11-0.16 | 0.01 | 0.01 | 0.14-0.19 | 0.67-0.71 |
| $\beta_i$ | 0.18-0.21 | 0.66-0.83 | 1.28-1.75 | 1.28-1.75 | 0.69-0.97 | 2.85-2.88 | 2.88-2.98 |
| $E_A^{(i)}$ (kV/mm) | 19.4 | 20.3 | 22.0 | 22.0 | 21.4 | 22.2 | 22.2 |
| $2P_s = 0.8\,\text{C/m}^2$, $\Delta S_{max} = -0.795\%$, $S_E = 6.6 \times 10^{-10}$ C/N, $\tau_0 = 0.7 \times 10^{-11}$ s ||||||||

The theory, in principle, properly captures the time dependence of both the polarization and strain responses for all considered field values. Fair quality of the polarization and strain description at lower fields may be related to the violation of the Merz law, which was established before for some ceramics in certain field regions [26,60,61]. A quasi-linear behavior of polarization and strain on the logarithmic scale at the final switching stage, not captured by the present theory, is known to originate from the statistical distribution of switching times [57,60] and can, in principle, be described by accounting of the latter [27].

Analysis of the field dependence of the fitting parameters reveals the following traits. Within the accuracy of determination, the fractions of different switching processes are approximately constant, see Fig. 6(a). The largest fraction corresponds to 180° processes (0.7), while the first 71°-switching process is characterized by the fraction $\eta_1$ of about 0.3. In contrast, fraction of the first 109°-events, $\eta_2 < 0.01$, remains virtually negligible. After the first step, the switching 71° process splits up in two – 71°- and 109°– possible events, characterized, respectively, by the fractions $\eta_{11}$ and $\eta_{12}$, making together $\eta_1$. The activation field for the first 71°- process, $E_A^{(1)}$, is the smallest of all and, particularly, smaller than that of the first 109°-process, $E_A^{(2)}$. The final 71° switching events over paths C-D and E-D in Fig. 1(a) are thermodynamically equivalent and thus the activation fields $E_A^{(111)}$ and $E_A^{(21)}$ coincide.

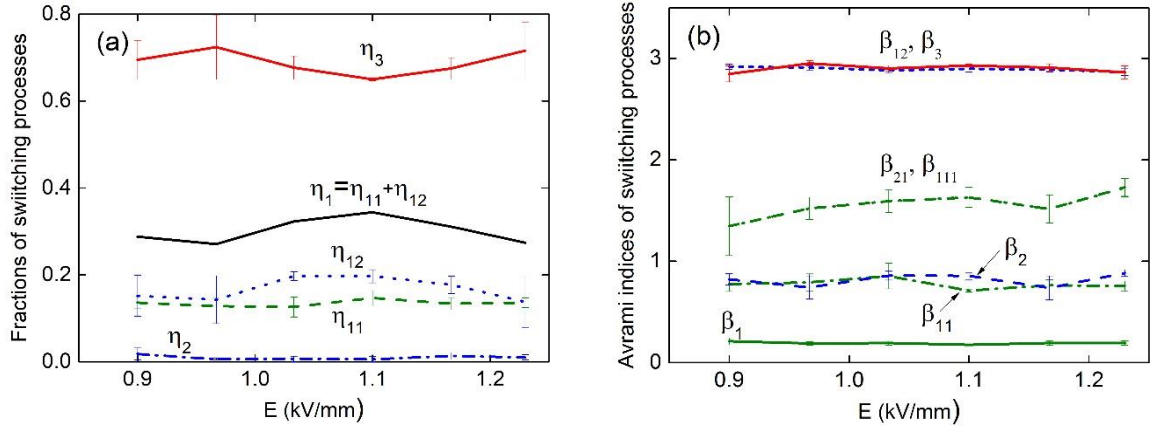

Fig. 6. Variation of (a) the fractions $\eta_i$ of the switching processes, depicted in Fig. 1(a), and (b) of the respective Avrami indices $\beta_i$ with the increasing applied field. Note the pairwise coincidence, within the inaccuracy of fitting, of the indices $\beta_{12}$ and $\beta_3$, $\beta_{21}$ and $\beta_{111}$, $\beta_2$ and $\beta_{11}$.

The Avrami indices of the first and the second 71°, second 109° and 180° switching are almost constant for all field values and correspond to $\beta_1 \cong 0.19$, $\beta_{11} \cong 0.7$, $\beta_{12} \cong 3$ and $\beta_3 \cong 3$, respectively (Fig. 6(b)). The indices of both equivalent final 71°switching events ($\beta_{21}, \beta_{111}$) are identical and increase with the field gradually from 1.28 to 1.75; however, this variation may be apparent considering the inaccuracy of their determination. In terms of the Ishibashi and Takagi classification [16], the switching events with $\beta_i < 1$ may only proceed according to the regime II from a number of latent nucleation sites being exhausted in the fast initial switching phase. This switching process seems to be facilitated by residual stresses in the initial highly polarized state, as was proposed in Ref. [19].

The most meaningful feature, revealed by a comparison of parameters of different processes, is the coincidence of the activation fields, $E_A^{(3)} = E_A^{(12)} = 22.2 \text{kV/mm}$, (and, thus of the respective switching times) and the respective Avrami indices, $\beta_3 = \beta_{12} \cong 3$, for the 180° (path A-D)- and the second 109°-processes (path B-D) in the 71°-109° succession. These activation fields are also not much different from that of the third 71°-process (path C-D). This is reminiscent of the recent results for the tetragonal PZT ceramics [26,27], where the kinetic switching characteristics for the second 90°- and the 180°-processes appeared to be identical. In the tetragonal case, a hypothetical coherent 90°-switching process suggested by Arlt [62], which may mimic the 180°-switching, could be suggested as the explanation of this coincidence. Though such coherent 109°- or 71°-switching processes have not been directly observed in experiments, the similarities in switching behaviour of tetragonal and rhombohedral ceramics make it reasonable to look for such a phenomenon. In agreement with this hypothesis, the CSA model suggested by Li and Rajapakse [31] assumed that two or three types of non-180° switching events occur simultaneously. We note also that sequential 90°-processes were reported to dominate the response of tetragonal ferroelectrics [63].

*C. Hypothetical coherent non-180° switching processes in rhombohedral ferroelectrics*

How may a scenario of coherent non-180°-processes in rhombohedral materials, similar to that suggested by Arlt for tetragonal ferroelectrics [62], look like? To answer this question let us consider possible orientations of 109°- and 71°-domain walls in a rhombohedral perovskite with a cubic parent phase [64]. Fig. 7 shows examples of mechanically compatible and electrically neutral 109°- and 71°-domain walls taking, respectively, {001} and {101} planes.

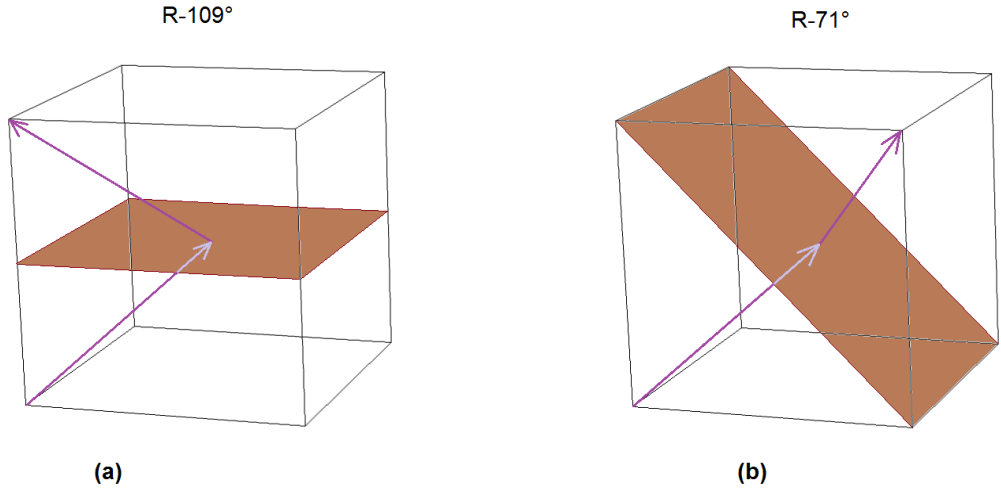

Fig. 7. Possible orientations of charge neutral and elastically compatible (a) 109°- and (b) 71°-domain walls in a rhombohedral ferroelectric [64].

Similarly to the proposal by Arlt [62], one can imagine a set of equally spaced parallel 109°-domain walls (Fig. 8(a)) separating domains whose polarizations are different by the sign of $x$- and $y$-components, as in Fig. 7(a). The direction of the mean polarization of this domain system $[001]$ is along the positive $z$-direction and normal to the domain planes. Applying electric field opposite to the mean polarization may create a domain wall in {110}-plane propagating in the negative $x$-direction, as is shown in Fig. 8(b) and (c). At this domain wall, only $z$-component of the polarization changes its sign, thus it presents a 71°-domain wall. Propagation of this wall across the stack of domains provides coherent 180°-switching of the mean polarization to the field direction ($[00\bar{1}]$) and does not change the macroscopic strain averaged over the domain stack. This process might mimic a 180°-switching of individual domains assumed by Eq. (2). To summarize this Section, the multi-step stochastic mechanism model is able to satisfactorily describe the time-dependent polarization and strain responses of rhombohedral ferroelectrics and reveal the fractions of different switching processes and their characteristics. However, the interpretation of the apparent process fractions remains questionable. The fractions of ferroelastically-active processes, $\eta_{11}, \eta_{12}$ and $\eta_2$ seem to be reliable. However, the fraction $\eta_3$ representing ferroelastically inactive processes may include direct 180° polarization reversals as well as coherent collective non-180° processes, as exemplarily presented in Fig. 8. This may explain the dominating apparent fraction of 180°-switching events in contrast to the expected highest energy barrier and thus the least likelihood of 180°-switching from the LGD-analysis. This hypothesis is supported by similar values of activation fields for the second 109°- after the first 71°-, the second 71°- after the first 109°-, the third 71°- and the 180°- switching events obtained from the fitting of experimental data.

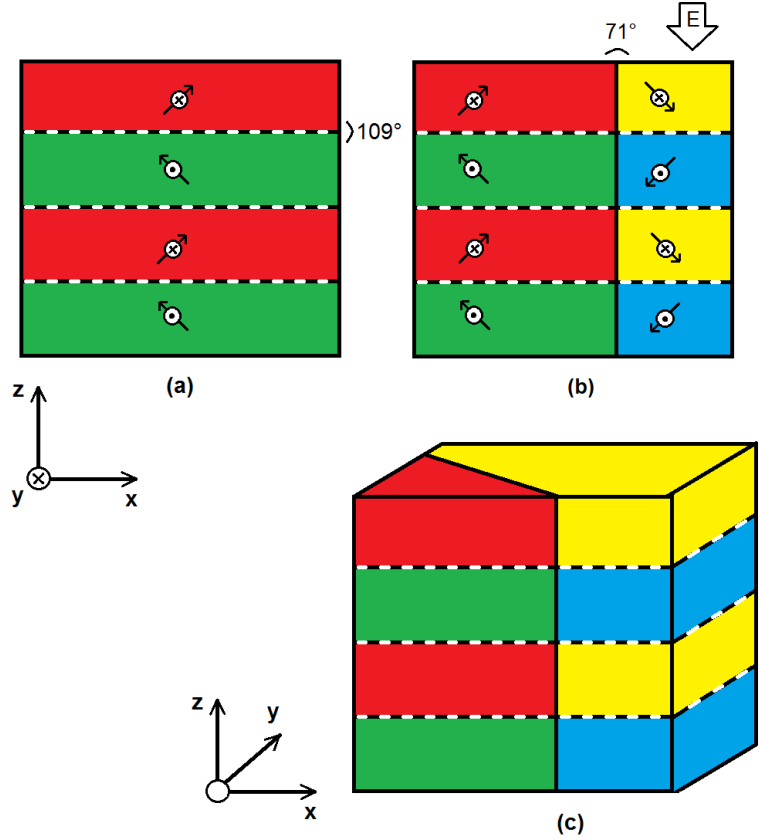

Fig. 8. (a) Array of equidistant 109°-domain walls (shown as dashed lines) in a rhombohedral ferroelectric with a mean polarization in $[001]$-direction. (b) A field-driven 71°-domain wall (shown as a solid line) sweeping through the domain system from right to left and switching the mean polarization to $[00\bar{1}]$-direction, thus providing macroscopic polarization reversal. (c) The 3D-sketch of the domain array to indicate the 71°-domain wall orientation.

## 5. Conclusions

Macroscopic electromechanical response of polycrystalline ferroelectric/ferroelastics can only be understood when the analysis includes sequential polarization rotation events. Consideration of consecutive switching events together with parallel ones opened for the first time a possibility to describe the dynamics of both electrical and mechanical response of ferroelectrics within one stochastic model. In the current work, the multi-step stochastic mechanism model of polarization switching [26] was advanced to investigate rhombohedral ferroelectrics, which allow sequential and parallel 71°-, 109°- and 180°- switching events, important for both fundamental understanding of the behaviour of these materials and for their applications.

Successful application of the model to recent simultaneous measurements of polarization and strain response of rhombohedral ferroelectric PZT ceramics over a wide time window [29] allowed detangling the individual switching events and quantifying their fractions, activation fields and Avrami indices, related to the dimensionality of the nucleated reversed domains. The fraction of ferroelastically active processes was established to be about 0.3 and dominated by the processes starting with 71°-switching events. The analysis of the likelihood of different switching paths by means of the LGD theory revealed that the prevalence of the first 71°- or the first 109°-rotations may depend on the residual stresses in the initial poled state. The dominating fraction of the ferroelastically inactive 180°-processes (about 0.7), obtained by means of the stochastic model, however, appeared unexpectedly large, considering that the LGD energy barrier for this process is the highest of all switching paths. In addition, the similarity of the activation fields of the 180°- and the final 109°- and 71°- switching events, resulting from the stochastic model analysis, raises a question on the genuine nature of the 180°-switching. These processes, contributing to the polarization variation but not contributing to the spontaneous strain variation, are reminiscent of the coherent processes in tetragonal materials, which do not develop macroscopically observable strain [62]. Such observations suggest a hypothesis that the response of rhombohedral materials is dominated by coherent non-180°- switching events, which mimic the 180°- switching processes.

**Acknowledgements**

This work was supported by the Deutsche Forschungsgemeinschaft (DFG) Grants Nos. 270195408 (GE 1171/7-1 and KO 5100/1-1) and GR 479/2. JED acknowledges financial support from the Australian Research Council Discovery Projects DP120103968 and DP130100415. JS acknowledges the support of the Feodor Lynen Research Fellowship Program of the Alexander von Humboldt Foundation. Dr. H. Kungl is acknowledged for the preparation of the sample and Prof. K. Albe and Dr. Y. Lysogorskiy for fruitful discussions.